\documentclass[prl,aps,showpacs,twocolumn,preprintnumbers,nofootinbib,amsmath,amssymb,superscriptaddress,longbibliography]{revtex4-2}

\usepackage{graphicx}

\usepackage[normalem]{ulem}
\usepackage[utf8]{inputenc}
\usepackage[T1]{fontenc}
\usepackage{mathptmx}
\usepackage{etoolbox}
\usepackage{makecell}
\usepackage{hyperref}

\draft 
\usepackage{xcolor}         

\begin{document}

\title{A scalable non-superconducting tunnel junction technology}

\author{J. Luomahaara}
\thanks{These authors contributed equally to this work.}
\affiliation{VTT Technical Research Centre of Finland Ltd, P.O. Box 1000, FI-02044 VTT Espoo, Finland}

\author{K. Razas}
\thanks{These authors contributed equally to this work.}
\affiliation{Department of Physics, University of Basel, Klingelbergstrasse 82, CH-4056, Basel, Switzerland}

\author{O. Sharifi Sedeh}
\thanks{These authors contributed equally to this work.}
\affiliation{Department of Physics, University of Basel, Klingelbergstrasse 82, CH-4056, Basel, Switzerland}

\author{R. P. Loreto}
\thanks{These authors contributed equally to this work.}

\author{J. S. Lehtinen}

\affiliation{VTT Technical Research Centre of Finland Ltd, P.O. Box 1000, FI-02044 VTT Espoo, Finland}

\author{M. Xu}
\affiliation{Department of Physics, University of Basel, Klingelbergstrasse 82, CH-4056, Basel, Switzerland}

\author{A. A. Cotten}
\affiliation{Department of Physics, University of Basel, Klingelbergstrasse 82, CH-4056, Basel, Switzerland}

\author{A. Tarascio}
\affiliation{Department of Physics, University of Basel, Klingelbergstrasse 82, CH-4056, Basel, Switzerland}

\author{P. Müller}
\affiliation{Department of Physics, University of Basel, Klingelbergstrasse 82, CH-4056, Basel, Switzerland}

\author{N. Yurttagül}
\author{L. Lehtisyrjä}
\author{L. Grönberg}
\affiliation{VTT Technical Research Centre of Finland Ltd, P.O. Box 1000, FI-02044 VTT Espoo, Finland}

\author{C. P. Scheller}
\affiliation{Department of Physics, University of Basel, Klingelbergstrasse 82, CH-4056, Basel, Switzerland}

\author{J. R. Prance}

\author{M. D. Thompson}

\author{R. P. Haley}
\affiliation{Department of Physics, Lancaster University, United Kingdom}

\author{M. Prunnila}
\email[]{Mika.Prunnila@vtt.fi}
\affiliation{VTT Technical Research Centre of Finland Ltd, P.O. Box 1000, FI-02044 VTT Espoo, Finland}

\author{D. M. Zumbühl}
\email[]{dominik.zumbuhl@unibas.ch}
\affiliation{Department of Physics, University of Basel, Klingelbergstrasse 82, CH-4056, Basel, Switzerland}

\date{\today}


\begin{abstract}
 Tunnel junctions are one of the key elements of chip-scale microsystems serving various technologies from classical microelectronics to quantum information. Aluminium and its oxide (AlOx) have dominated cryogenic tunnel junction technology for decades due to the high quality of AlOx barriers and Al superconducting properties below 1.2 K. However, many applications require non-superconducting junctions, either standalone or in combination with superconducting technology, motivating efforts to suppress Al superconductivity through magnetic fields, doping, or proximity effects—approaches that so far suffered from integration compatibility and scalability issues. Here, we present a CMOS-compatible normal-metal tunnel junction technology based on TiW alloy and AlOx barriers. We demonstrate wafer-scale fabrication of TiW/Al-AlOx/TiW junctions and validate their performance in Coulomb blockade thermometers operating down to 20 mK, confirming robust normal-state behavior. This TiW-based architecture offers a scalable solution for non-superconducting tunnel junctions across a broad temperature range, enabling integration into advanced cryogenic, quantum and nanoelectronic chip-level systems.
\end{abstract}

\maketitle
Tunnel junctions enable precise control of charge transport at the nanoscale and play a fundamental role in modern microelectronics. They are implemented in a wide range of components and architectures, built from combinations of normal metals, superconductors, dielectrics, and ferroelectric and ferromagnetic materials. As a result, tunnel junctions underpin diverse technologies, including memories \cite{ike, sha2}, spintronics \cite{mac}, neuromorphic computing \cite{cov}, and quantum information processing \cite{kra}.
Metallic junctions can either turn superconducting below the critical temperature or remain in the normal-metal state down to the lowest achievable chip temperatures. Normal-metal tunnel junctions appear for example in calibrated noise sources \cite{mal} and primary thermometers like Coulomb blockade thermometers (CBT) \cite{pek} and shot noise thermometers \cite{spi1, spi2}. Superconducting junctions -- Josephson junctions -- are essential building blocks for superconducting qubits \cite{kra}, sensitive magnetometers \cite{cla}, and high-speed classical logic circuits \cite{bai}. Hybrid junctions joining a normal-metal with a  superconductor through an insulator are used in electron thermometers \cite{fes2}, radiation detectors \cite{nah} and electronic coolers \cite{myk, hat}.

For metallic tunnel junction devices operated at cryogenic temperatures, like superconducting qubits \cite{kra} that are proliferating widely due to strongly growing quantum technology activities, aluminium  and its oxide has been the prevalent workhorse material  since 1980s \cite{gia, cla}. This comes  with a  blessing and a curse: the thermal oxide of Al, AlOx, is a naturally forming  tunnel barrier of excellent quality and Al becomes an almost ideal superconductor below the critical temperature of $\sim$ 1.2 K. While superconductivity can be beneficial and desired, also non-superconducting tunnel junctions are generally needed, either as a standalone technology or for integration with other elements and circuits – like in many of the examples discussed above \cite{pek, spi1, spi2, mal, fes2, nah, myk, hat}.

The high quality of the AlOx tunnel barrier on top of Al has inspired suppression of  the Al superconductivity for the creation of normal-metal junctions. Applying a magnetic field is a brute force way of destroying the superconductivity, but it lacks universal applicability, and more intimate and scalable methods have been searched for over the years. Such suppression strategies have involved, for example, Mn doping \cite{rug} or employing reverse proximity effects from non-superconducting metals, such as Cu \cite{kos}. Although these approaches have provided significant reduction in Al critical temperature, these materials — as well as noble metals— are generally incompatible with many nanoelectronic circuits and technologies, like those incorporating superconductors. This incompatibility arises, for example, from their magnetic properties, detrimental diffusion behaviour, or the inherent limitations in fabrication reliability and scalability.

In this work, we report on a normal-metal tunnel junction technology that is based on a TiW alloy that is a typical CMOS compatible diffusion barrier and adhesion layer \cite{fra} and has also been employed as a resistor material in cryogenic circuits \cite{kiv}. We utilize a TiW/Al-AlOx/TiW material stack to benefit from the TiW normal-state conduction and the high quality of AlOx as a tunnel barrier. We show high yield fabrication of the tunnel junction elements on the wafer scale and a practical thermometry use-case with TiW-based CBTs down to 20 mK demonstrating the normal-state operation of the stack at zero magnetic field. Our TiW-based approach introduces a normal-metal tunnel junction building block that can be utilized in a variety of applications calling for scalable non-superconducting tunnel junctions in a broad temperature range down to deep cryogenic regime.

\begin{figure*}[htpb]
\centering
\includegraphics[width=0.99\linewidth]{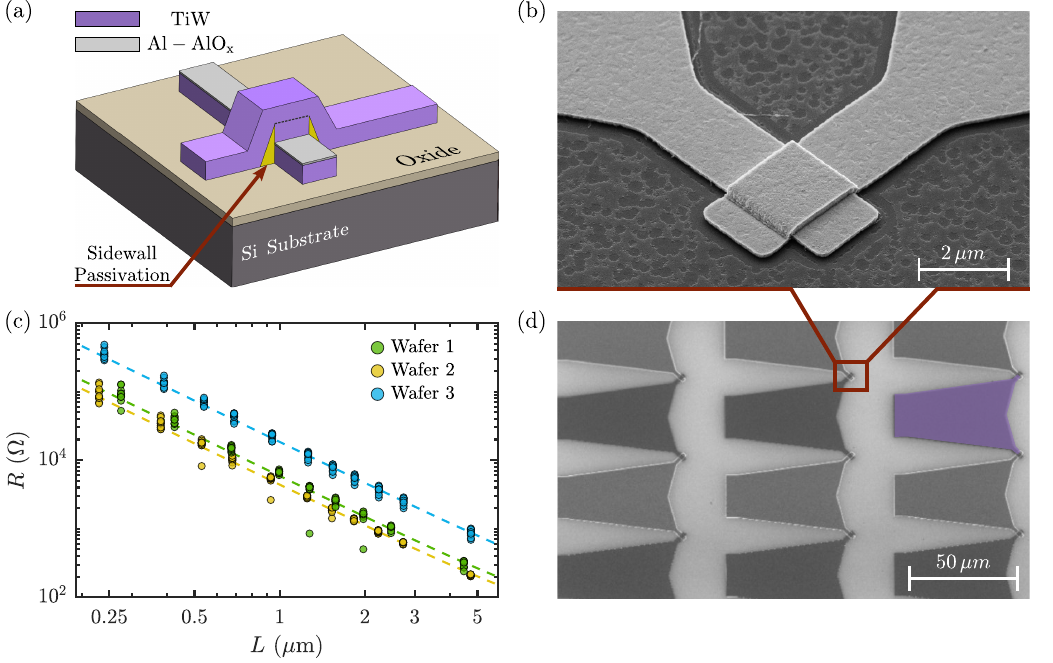}
\caption{\textbf{a}, A cross-sectional illustration of the fabrication process. \textbf{b}, A tilted SEM image showing a cross-type 2.1 $\mu$m junction after patterning the upper metal layer. \textbf{c}, Room-temperature probing data as a function of a realized junction width $L=d_\mathrm{J}-\mathrm{LWR}$ (see main text) over three wafers with different tunnel resistance levels as characterized by specific resistivity $\rho_\mathrm{t}$ (Methods). Markers denote the measurement data, and lines are fits to the data as specified in the main text. \textbf{d}, An SEM image of the CBT. Large islands between the junctions were included in the designs, leaving space for possible metal electroplating in future fabrication processes. The purple colour indicates an island and the red outline marks the junction. The different shades of gray represent the wiring layer and the base electrode of the trilayer \cite{gro}, which alternate within the array.}
\label{device}
\end{figure*}

To create a wafer-scale non-superconducting tunnel junction technology, we keep relying on the excellent properties of Al but suppress the superconductivity down to low temperatures with a proximate CMOS-compatible pair-breaking metal layer. To implement this, we modify a recently developed tunnel junction process \cite{gro} by introducing a proximate TiW alloy (purple) in a TiW/Al-AlO$_{\rm x}$/TiW trilayer  junction stack (Fig.~\ref{device}a). The junction technology has been previously applied successfully to numerous superconducting components such as Josephson parametric amplifiers \cite{sim, per}, microcoolers \cite{hat} and qubits \cite{set}. Prior to this work, wafer-scale technology for CBTs was based on standard Al ex-situ junctions requiring a magnetic field to keep it in the normal-state \cite{pru, bra}. An important feature of the technology is the sidewall passivation (yellow) that prevents short circuits in the junction and enables junction definition in a compact cross-type configuration.  Fig.~\ref{device}b shows an SEM image of a completed junction.

To validate the fabrication process, several test wafers are designed and manufactured, incorporating both single tunnel junction devices and their arrays configured as CBTs. Properties of the tunnel junctions, such as specific resistivity $\rho_{\mathrm{t}}$ and the fabrication process-induced line width reduction (LWR), are extracted by measuring and fitting the resistance using the relationship $R = \rho_{\mathrm{t}}/(d_{\mathrm{J}}-\mathrm{LWR})^{2}$, where $d_\mathrm{J}$ is the design value for junction width. Fig.~\ref{device}c shows the room-temperature resistance $R$ of individual tunnel junctions as a function of their effective width $L= d_{\mathrm{J}}-\mathrm{LWR}$ measured across three distinct wafers. The resistances obtained demonstrate excellent agreement with the fitted curves and our previous reports for this type of superconducting tunnel junctions \cite{gro}, confirming the scalability of the process down to sub-micron dimensions at the wafer level (Methods show the whole fabrication procedure in more detail).

\begin{figure*}[htpb]
\centering
\includegraphics[width=0.95\linewidth]{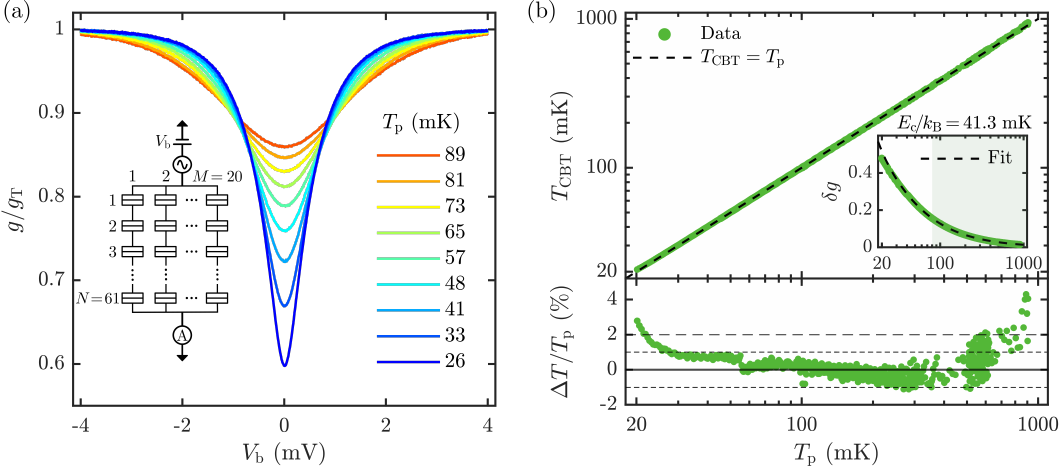}

\caption{\textbf{a}, Normalised conductance $g/g_{\mathrm{T}}$ of the CBT as a function of bias voltage $V_{\mathrm{b}}$ at different temperatures. Inset: measurement scheme of the CBT. \textbf{b}, Comparison of the sample holder phonon temperature $T_{\rm p}$ to the electron temperature $T_{\rm CBT}$ measured by the secondary thermometry mode. Inset: differential conductance dip $\delta g$ versus $T_{\rm p}$. The dashed curve is the third-order polynomial fit, with fit domain indicated as shaded background, delivering $E_{\rm c}$ with fit error $<0.5\%$. Lower panel: the relative deviation $\Delta T=T_{\rm CBT}-T_{\rm p}$ normalised by $T_{\rm p}$, remaining mostly below $1\%$ (see main text). The small glitch at $T_{\rm p}\sim55~\rm mK$ is an artefact due to a change in the point-spacing of the $T_{\rm p}$ calibration.}
\label{fig:measurement}
\end{figure*}

CBTs provide a well-established technology for measuring electron temperature \cite{fes,yur,mes2,cas,pal} with demonstrations extending from sub-mK \cite{sam} to 60~K \cite{mes}. Such CBTs work only if the junction electrodes are normal-conducting, hence they offer an ideal platform for confirming the absence of superconductivity while providing accurate on-chip thermometry within the same device. Therefore, to characterise the new normal-metal tunnel junction technology at low temperatures, we assembled junctions into an $N \times M$ array where $N$ is the number of junctions in series with islands in between, and $M$ denotes the number of chains connected in parallel (Figs.~\ref{device}d and ~\ref{fig:measurement}a).

The basic functionality of a CBT is based on the Coulomb charging energy $E_{\rm c}=\frac{N-1}{N}\frac{e^2}{C}$, representing the energy required to add an extra electron to an island with capacitance $C$ positioned in between each two tunnel junctions. The necessity to overcome this energy leads to a suppression of the electrical current through the array around zero voltage bias -- the Coulomb blockade effect \cite{ave,gra}. As a consequence, the differential conductance $g=\mathrm{d}I/\mathrm{d}V$ also exhibits a dip at low voltage bias $V_{\rm b}$, which becomes more pronounced as the CBT temperature $T_{\rm CBT}$ approaches the temperature scale associated with the charging energy $E_{\rm c}/k_{\rm B}$, as seen in  Fig.~\ref{fig:measurement}a. In principle, the width of this dip is directly proportional to temperature, making CBTs primary thermometers \cite{pek,sha}.
In practice, applying a finite voltage bias leads to ohmic dissipation and overheating effects, particularly severe at low millikelvin temperatures. Instead, the normalised depth $\delta g=(g_{\rm T}-g_0)/g_{\rm T}$ of the zero-bias conductance dip $g_0$ with respect to the high-bias conductance $g_{\rm T}$ is a better measure of temperature since it does not suffer from such heating issues. To convert $\delta g$ into a temperature, the charging energy is required. This can be extracted by calibrating the CBT against another thermometer, thus making this mode of operation a secondary thermometer \cite{cas,fes,sha,yur2}.

Observing the typical CBT bias traces shown in Fig.~\ref{fig:measurement}a suggests that the junctions remain normal-conducting down to the lowest temperatures presented. The signatures of superconductivity would further push down the conductance at low bias (in the dip region) below the normal-conducting CBT behaviour due to a vanishing density of states in an opening superconducting gap, eventually reaching very low or zero conductance when temperature is far below the superconducting transition \cite{kos,cha,nev}. Further, the superconducting gap structure with large density of states just above the energy gap would typically lead to an overshoot of the differential conductance with bias close to the superconducting gap \cite{tinkham}. Neither of these signatures can be seen in the data here. Instead, a continuously evolving smooth temperature dependence is observed, presenting the first evidence of non-superconducting tunnel junctions.

\begin{figure*}[htpb]
\centering
\includegraphics[width=0.95\linewidth]{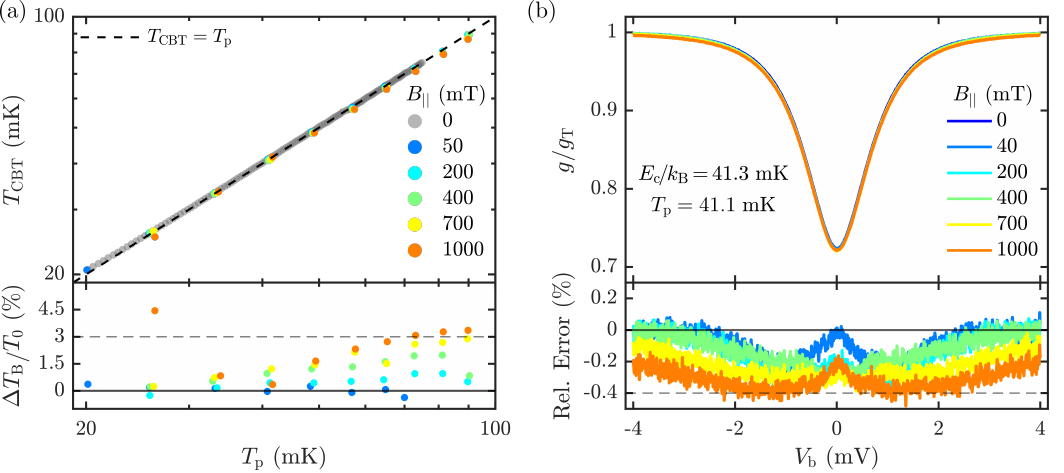}
\caption{\textbf{a}, CBT electron temperature $T_{\rm CBT}$ with respect to $T_{\rm p}$ measured at different parallel magnetic fields. Each temperature step was held for $\sim2$  hours to ensure proper thermalization. The lower panel displays the relative deviation between field and zero-field measurement. The difference is defined as $\Delta T_{\rm B}=  T_{0} - T(B_{||})$, where $T_{0}$ is the temperature extracted at zero field, serving as the normalisation reference. $T_{\rm CBT}$ was determined using MCMC simulations parametrised by prior zero-field measurements. \textbf{b}, Normalised differential conductance $g/g_{\mathrm{T}}$ of the CBT as a function of bias voltage $V_{\mathrm{b}}$ at different parallel fields and a fixed temperature $T_{\rm p} = 41$~mK. An additional panel shows the discrepancy introduced by the magnetic field, where the relative error is defined by normalising the field traces against the zero-field reference measurement.}
\label{fig:supp}
\end{figure*}

Next, we operate the CBT as a precision thermometer across a large temperature range down to 20~mK.
To calibrate the CBT and retrieve its temperature, we measure the depth of the zero-bias conductance dip $\delta g$ as a function of the phonon temperature $T_{\rm p}$ obtained using a factory-calibrated ruthenium-oxide thermometer inside the sample holder. We fit this data in the high-temperature regime (green-shaded area in the inset of Fig.~\ref{fig:measurement}b) using a third-order approximation of the master equation \cite{fes,far}. In this range, the two thermometers are assumed to be thermalised, allowing us to estimate the charging energy as a fit coefficient. The resulting charging energy for this device turns out relatively high, $E_{\rm c}/k_{\rm B}=41.3\pm0.2~\rm mK$, and pushes the low temperature-end of the measurement outside the validity range of the third-order approximation, as the conversion error already exceeds $1.2$\% when $k_{\rm B}T_{\rm CBT}  \sim 0.8 E_{\mathrm{c}}$ \cite{fes,sam,sha,yur}. Below this limit, offset charges present on the islands can start to significantly affect the conductance and thus the temperature reading -- a factor that was negligible at higher temperatures. To include these effects, Markov chain Monte Carlo (MCMC) simulations can be used \cite{yur, sha} to extend the thermometry further with relatively low uncertainty. In this study, MCMC simulations estimate a maximum thermometry error of 0.8\% at low temperatures, which diminishes towards higher temperatures (Methods; Supplementary Fig.~\ref{fig:supp3}).

Following the above method, the temperature can be extracted reliably from the conductance dip $\delta g$, as shown in Fig.~\ref{fig:measurement}b. The correspondence between the CBT and sample holder temperatures is so strong that the two traces fully coincide almost indistinguishably on the log-log plot extending over almost two orders of magnitude in temperature from 20~mK to 1~K. To inspect the agreement in more detail, we plot the relative deviation $\Delta T=T_{\rm CBT}-T_{\rm p}$ normalised by $T_{\rm p}$, as shown in the lower panel. The discrepancy is less than $1\%$ over most of the range, except for the highest temperatures, where the dip $\delta g$ becomes very small (comparable to the electronic noise) and subject to possible temperature dependence of the high-bias differential conductance $g_{\rm T}$ \cite{xu}. On the other side of the temperature range, below 30~mK, $T_{\rm CBT}$ is seen to rise slightly above $T_{\rm p}$. This frequently observed effect is well understood and is attributed to the dominant electron-phonon cooling mechanism within the thin metallic islands of the CBT, typically scaling as $T_{\rm p}^5$ and thus becoming insufficient at low temperatures \cite{mes3,sha,gia,cas,sch}. Furthermore, if the tunnel junctions were indeed weakly superconducting, the suppression $\delta g$ would lead to lower values and consequently, the inferred CBT temperatures $T_{\rm CBT}$ would decrease below $T_{\rm p}$ -- contrary to the experimental observations.

To further validate the absence of superconductivity, we repeat the $\delta g$ measurements and determine $T_{\rm CBT}$ at various magnetic fields $B_{||}$ parallel to the substrate, as shown in Fig.~\ref{fig:supp}a. The results demonstrate that the CBT is almost completely insensitive to the external magnetic field, which not only confirms the normal conductivity of the tunnel junctions but also highlights the reliability of the thermometry at high magnetic fields \cite{pek3, pek4}. The field-induced deviation is very small and is more clearly identified in the lower panel plot. The difference in $T_{\rm CBT}$ remains within $\lesssim 3\%$ throughout the entire 20--100~mK range, except for the data measured at a high 1~T parallel field. The robustness of the CBT is further supported by Fig.~\ref{fig:supp}b, where the effects of the magnetic field on the differential conductance traces are largely negligible. A slight field dependence of $g_{\rm T}$ is observed, which is likely responsible for the slight deviation at high-field shown in Fig.~\ref{fig:supp}a.

In conclusion, we have developed a novel wafer-scale process for fabricating normal-metal tunnel junctions by introducing TiW as an electrode material. This electrode material has no magnetic effects and is CMOS-compatible; therefore, such tunnel junctions can be scaled and coexist with both normal and superconducting circuits. To demonstrate the approach, we fabricate an array of tunnel junctions forming a Coulomb blockade thermometer and present its operation at zero magnetic field and temperatures as low as 20~mK. Introducing the magnetic field into the measurements shows only a negligible effect, confirming the complete absence of superconductivity all the way to the lowest temperatures reported for normal-insulator-normal junctions, thus overcoming the previous limitations of zero-field tunnel junction technology. We note also that the presented junction technology shows good yield over the wafer, as verified by the room-temperature probing data of individual junctions and cryogenic characterization of multiple CBTs (see Supplementary Information).

Ongoing developments aim to extend the normal-conducting regime into the microkelvin temperature range by ensuring effective thermalisation of the metallic islands while maintaining sufficient suppression of superconductivity. Further, hybrid junctions built with this new technology could be explored for thermometry and studying the material and junction properties with subgap spectroscopy. Finally, the fact that CBTs now work in the absence of a magnetic field will help to spread this method for easy and convenient thermometry in low temperature experiments.



%

\newpage

\section{Methods}

\subsection{Preparation of samples}

The fabrication process begins with a dielectric layer formed by thermal deposition of SiO$_{2}$, followed by Atomic Layer Deposited (ALD) layer of AlO$_{x}$ on a silicon substrate. Next, a normal-metal trilayer stack (TiW / Al-AlO$_{x}$ / TiW) is sputtered \emph{in-situ} without breaking the vacuum. The first two layers, TiW and Al, are sputtered to thicknesses of 100 nm and $\sim$7 nm, respectively, with the Al layer subsequently oxidised to form a tunnel barrier. The oxygen pressure and oxidation time define the junction specific resistivity.  Another 100-nm-thick TiW layer completes the trilayer. The trilayer is patterned using UV lithography and etched using reactive ion etching (RIE). Then, a SiO$_{2}$ layer is deposited using plasma-enhanced chemical vapour deposition (PECVD), followed by anisotropic plasma etching that leaves the spacer structures on the sidewalls of the trilayer stack. Finally, a TiW wiring layer of thickness 100 nm is deposited, patterned, and plasma-etched completing the cross-type junction structure \cite{gro}.

The nominal width of the test junctions are varied from 500~nm to 5~$\mu$m and the junctions are distributed across the 150~mm wafers. For each width, twelve junctions are measured at room temperature, totalling 132 measured junctions per wafer (see Fig.~\ref{device}c). The obtained junction specific resistivities $\rho_\mathrm{t}$ are 6900~$\Omega \mathrm{ \mu m^2}$, 4800~$\Omega \mathrm{ \mu m^2}$ and 21000~$\Omega \mathrm{ \mu m^2}$ for wafers 1, 2 and 3, respectively, while the extracted LWR values are 0.53~$\mu$m, 0.27~$\mu$m and 0.26~$\mu$m, which are consistent with our previous reports for this type of superconducting tunnel junction technology \cite{gro}.

To characterise the junction technology at low temperatures, a CBT design with a large 61$\times$20 array of tunnel junctions is adopted for experiments. The number of junctions in series increases the tolerance against the voltage noise and suppresses the impact of the electromagnetic environment on measurement accuracy, while introducing many identical rows in parallel lowers the impedance, enabling a sufficiently large current through the device. Additionally, having more parallel chains further enhances the CBT accuracy for applications beyond the universal regime as the uncertainty introduced by the random charge offsets in the thermalisation islands is greatly reduced \cite{yur}. Figures \ref{device}(b) and (d) illustrate the fabricated devices with each junction having a nominal width $d_\mathrm{J} =$ 0.95 $\rm{\mu} \rm m$ ($\sim$ 0.4 $\rm{\mu}\rm m$ after linewidth reduction). Thus, the array results in a CBT with charging energy $E_{\rm c}/k_{\rm B} \sim 41.3 \pm 0.2$~mK. The charging energy sets the temperature range of thermometry, and operation across different temperature ranges requires varying the $E_{\rm c}$ and hence the area of the junctions. This is made possible by the established and precise control of junction dimensions demonstrated above.

\subsection{Low-temperature characterisation}

Devices from wafer 1 (Fig.~\ref{device}c) were chosen for cryogenic experiments. The measurements were performed in a Bluefors LD dilution refrigerator, equipped with a fast sample exchange (FSE) probe mechanism. The probe included a calibrated ruthenium oxide thermometer, which we assume accurately reflects the phonon temperature and is denoted as $T_{\mathrm{p}}$. The experimental data was obtained using standard lock-in techniques with current monitoring and 4 $\mu \rm V$ voltage bias excitation at 77 Hz frequency.

Fig.~\ref{fig:measurement}(a) shows the measurements of CBT's relative conductance $g/g_{\mathrm{T}}$ as a function of temperature in the range of $T_{\rm p}\approx26-90$~mK at zero magnetic field. Here, the temperature is swept from cold to warm in steps, where a waiting time of $\sim2$~h is used to ensure sufficient stabilisation of the refrigerator temperature and proper equilibration of the sample. Any possible magnetic field offsets are compensated to ensure zero-field condition.

In Fig.~\ref{fig:measurement}b, secondary thermometry operation of the CBT is illustrated. The temperature is swept from cold to warm in small steps, leaving time for thermalization. The procedure is repeated for multiple samples with identical design measured in different cooldowns (Supplementary Fig.~\ref{fig:supp2}). The results are qualitatively the same for all three samples and confirm that the tunnel junctions consistently remain normal-conducting all the way to 20~mK. All devices show similar resilience to the applied magnetic field, as previously discussed in the main text (Fig.~\ref{fig:supp}).

\subsection{Temperature conversion}

CBT electron temperature $T_{\rm CBT}$ is extracted from $\delta g$ as a function of probe temperature $T_{\rm p}$. The conversion from $\delta g$ to $T_{\rm CBT}$ is performed using both analytical approximations and numerical simulations, which introduce a quantifiable error margin.

At high temperatures $T_{\rm CBT}$ is calculated using the third-order approximation of the master equation \cite{far}
\begin{equation*}
\delta g = \frac{1}{6}\left(\frac{2E_{\rm c}}{k_{\rm B}T_{\rm CBT}}\right) - \frac{1}{60}\left(\frac{2E_{\rm c}}{k_{\rm B}T_{\rm CBT}}\right)^2 + \frac{1}{630}\left(\frac{2E_{\rm c}}{k_{\rm B}T_{\rm CBT}}\right)^3.
\end{equation*}
The expression above is applicable from high temperature regime down to $T_{\rm CBT} \sim 0.8 E_{\mathrm{c}}/k_{\rm B}$ \cite{sam,sha,yur}. Beyond this point, the error exceeds 1.2\% and increases rapidly when lowering the temperature further, hence it marks the end of the validity range of the approximation. For measurements in Fig.~\ref{fig:measurement}b, this method is only used down to $\sim45$~mK to avoid the readily increasing error.

Beyond the universal regime ($T_{\rm CBT} < 0.8 E_{\mathrm{c}}/k_{\rm B}$), the conductance is heavily affected by the offset charges induced on the island, thus $T_{\rm CBT}$ has to be obtained using MCMC numerical calculations that include these effects. MCMC simulations are performed with parameters specifically tuned to represent the investigated CBT and result in an uncertainty that remains within $<1$\% throughout the entire temperature range of this experiment. In particular at high temperatures, the resulting calibration agrees with the fridge thermometer $T_{\rm p}$ exactly. Precise details and limiting cases are discussed and illustrated in the Supplementary Information.

\section{Acknowledgements}
We thank J. Pekola, E. Praks and L. Wang for useful discussions.

The research was financially supported by the European Union’s Horizon RIA and EIC programmes under Grants No. 824109 European Microkelvin Platform (EMP), No. 101113086 SoCool, and No. 101113983 Qu-Pilot. We also acknowledge financial support of Research Council of Finland through project No. 350667 Femto, No. 336817 the QTF Centre of Excellence project and No. 374172 QMAT Centre of Excellence. Financial support was provided also by Business Finland through project CryoTherm and No. 128291 Quantum Technologies Industrial (QuTI) as well as by Technology Industries of Finland Centennial Foundation and Chips JU project Arctic No. 101139908.
Additional support came from the MSCA Cofund Action Quantum Science and Technologies at the European Campus (QUSTEC) under Grant No. 847471, the Swiss National Science Foundation (Grant No. 179024), the Swiss Nanoscience Institute, and the Georg H. Endress Foundation.
M.D.T. acknowledges financial support from the Royal Academy of Engineering (RF/201819/18/2).

The data supporting the plots of this paper will be available at a Zenodo repository.

\section{Contributions}
M.P. initiated the research. J.S.L, N.Y. and R.P.L. carried out the process development during which J.R.P., M.D.T and R.P.H. performed preliminary experiments to characterise the junction arrays at low temperatures. J.L. designed the devices with inputs from R.P.L., O.S.S. and M.P. Samples were fabricated by R.P.L. together with G.L. The room temperature probing of junctions was performed by L.L. and R.P.L. with J.L. analysing this data. K.R., O.S.S., M.X. and A.A.C. performed low temperature transport measurements of the developed CBTs and analysed the data shown in the manuscript, with A.T., P.M. and C.S. ensuring operation of the low temperature setup. K.R, J.L., M.P. and D.M.Z. wrote the manuscript with inputs from all authors. D.M.Z. supervised the low temperature experiments and M.P. the component development work. All authors contributed to the discussions and understanding of the results.






\section{Supplementary information}

\subsection{Reproducibility of experiments}

\begin{figure}[htpb]
\centering
\includegraphics[width=1\linewidth]{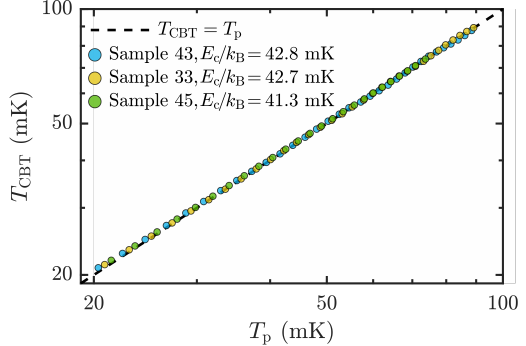}
\caption{Electron temperature of the CBT ($T_{\rm CBT}$) from different devices as a function of phonon temperature $T_{\rm p}$. Each sample is measured one-by-one in separate cooldown instances.}
\label{fig:supp2}
\end{figure}

The reproducibility of the junction technology and measurements is demonstrated in Fig.~\ref{fig:supp2}, where the same experiment is repeated for three different devices picked from different locations of wafer 1, each measured at a separate cooldown. Each CBT sample has an identical design as described in the Methods section. The converted temperatures $T_{\rm CBT}$ using MCMC simulations are consistent and overlap completely, hence only sparse data points are shown to improve clarity.

\subsection{Thermometry errors}

\begin{figure}[htpb]
\centering
\includegraphics[width=1\linewidth]{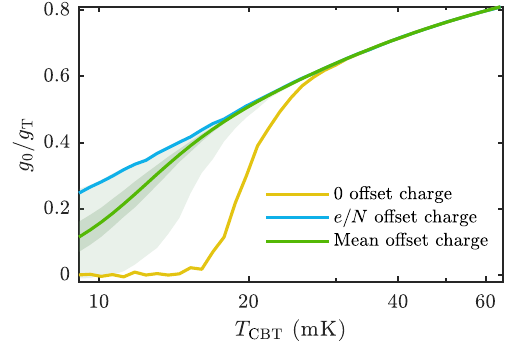}
\caption{MCMC simulations of temperature dependence of the zero-bias differential conductance for different offset charge configurations. The yellow and blue lines represent maximal and minimal Coulomb suppression, respectively. The wide shaded area is the range of simulation outcomes with random offset charge for a CBT with one row of junctions. The smaller shadow region is the uncertainty for a CBT with 20 parallel rows. The solid green line denotes the mean of the former region and is used to extract $T_{\rm CBT}$ from the measured differential conductance.}
\label{fig:supp4}
\end{figure}

Fig.~\ref{fig:supp4} shows the distribution of conductance curves occurring from the MCMC simulations. Each simulation cycle models a single chain of $N=61$ tunnel junctions and has a random offset charge per island. This yields a unique temperature dependence of the zero-bias conductance for every configuration. All configurations have offset charges chosen randomly in the range of $\pm e/2$, which result in a family of curves (wide green shaded area) positioned in between the two furthermost curves (yellow and blue). The yellow curve is simulated with no offset charge and represents the maximal Coulomb blockade case, while the blue curve is obtained when each island contains a fractional $e/N$ offset charge providing minimal suppression. All simulation cycles are then grouped by parts of 20 and averaged to represent the measured CBT with 20 chains connected in parallel. This averaging reduces the uncertainty induced by offset charges and provides a much narrower region of possible outcomes (narrow green shaded area). The mean value of this region, depicted as solid green curve, is used to extract the temperature in the experiment. The uncertainty here is small, and is expected to be so, as it is known that large number of junctions connected in series and more chains in parallel increase the accuracy of thermometry \cite{yur,sha}.

\begin{figure}[htpb]
\centering
\includegraphics[width=1\linewidth]{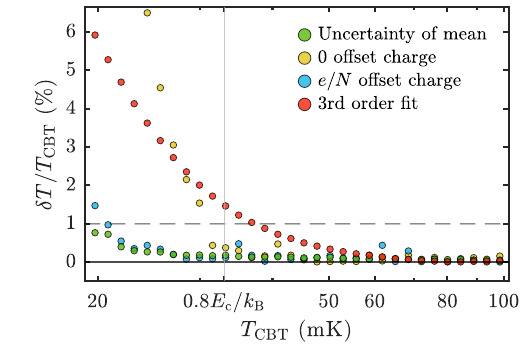}
\caption{Relative error in temperature reading ($\delta T/ T_{\rm CBT}$) using different conversion approaches as a function of $T_{\rm CBT}$. Here, the continuous green curve in Fig \ref{fig:supp4} is taken as the reference conversion curve. Error of the extreme limits are in yellow and blue dots, respectively. Red dots show the error of the third-order approximation of the master equation. The green dots show the uncertainty caused by offset charges on the MCMC mean conversion curve. The dashed line highlights the 1\% error range.}
\label{fig:supp3}
\end{figure}

Fig.~\ref{fig:supp3} highlights the importance of MCMC simulations compared to conventional methods. The uncertainty induced by simulations (green dots) is presented in terms of relative temperature difference with respect to the mean value discussed before. The third-order approximation of the master equation is included in the figure as red dots. At temperatures $T_{\rm CBT}>80$~mK, the resulting temperatures are identical and do not depend on the conversion method. Conversely, when $T_{\rm CBT}\leq0.8E_{\rm c}/k_{\rm B}$, the error in third-order approximation starts readily increasing hence the transition to MCMC simulations is essential.

\end{document}